\documentclass[10pt,a4paper]{article}

\usepackage[cm]{fullpage}

\usepackage{epsfig}
\usepackage[fleqn,sumlimits,intlimits]{amsmath}
\usepackage[english]{babel}
\usepackage{cite}

\usepackage[active]{srcltx}
\usepackage{slashed}
\usepackage{verbatim}
\usepackage{amsmath,latexsym}
\usepackage{amssymb}
\usepackage{hyperref}
\usepackage{IEEEtrantools}

\def\on#1#2{{\buildrel{\mkern2.5mu#1\mkern-2.5mu}\over{#2}}}

\newcommand{\beq}{\begin{equation}}
\newcommand{\eeq}{\end{equation}}

\newcommand{\myref}[1]{(\ref{#1})}

\makeatletter
\DeclareRobustCommand{\cev}[1]{%
  {\mathpalette\do@cev{#1}}%
}
\newcommand{\do@cev}[2]{%
  \vbox{\offinterlineskip
    \sbox\z@{$\m@th#1 x$}%
    \ialign{##\cr
      \hidewidth\reflectbox{$\m@th#1\vec{}\mkern4mu$}\hidewidth\cr
      \noalign{\kern-\ht\z@}
      $\m@th#1#2$\cr
    }%
  }%
}
\makeatother

\begin{document}

\title{A First Principles Derivation of Classical and Quantum Mechanics as the Natural Theories for Smooth Stochastic Paths}
\date{}
\vspace{-10pt}
\maketitle

\begin{center}

\vspace{-50pt}

{\sl Willem\ Westra}$\,^{a, b}$

\vspace{5pt}

{\footnotesize
$^a$~Owlin \\
Stadhouderskade 85, 1073AT Amsterdam, Netherlands.\\
{ email: willem@owlin.com}\\

\vspace{0pt}

$^b$~Department of Physics, University of Iceland,\\
Dunhaga 3, 107 Reykjavik, Iceland\\

}
\vspace{0pt}

\end{center}

\begin{abstract}
We derive the classical Hamilton-Jacobi equation from first principles as the natural description for smooth stochastic processes when one neglects stochastic velocity fluctuations. The Schr\"{o}dinger equation is shown to be the natural exact equation for describing smooth stochastic processes. 
In particular, processes with up to quadratic stochastic fluctuations are electromagnetically coupled quantum point particles. 
The stochastic derivation offers a clear geometric picture for Quantum Mechanics as a locally realistic hidden variable theory. 
While that sounds paradoxical, we show that Bell's formula for local realism is incomplete. 
If one includes smooth stochastic fluctuations for the hidden variables, local realism is preserved and quantum mechanics is obtained.
Quantum mechanics should therefore be viewed as a ``nondeterministic, non-Bell locally realistic hidden variable theory".
Since the description is simply a stochastic process, it should be relatively straightforward to create mesoscopic analogue systems that show all the hallmarks of Quantum Mechanics, including super-Bell correlations. In fact, any system that can be described by the linear time evolution of a density matrix is both a stochastic and a Quantum system from our point of view, since we show that the existence of a transition probability density directly implies the existence of a density matrix. Systems for which the stochastic degrees of freedom vary smoothly over time have quantum Hamiltonians with standard kinetic terms.
\end{abstract}
\vspace{-.4cm}
\section{Introduction}
\vspace{-.2cm}
For many years ,the parallels between quantum mechanics and stochastic processes have been discussed from various points of view \cite{Fenyes:1952, Takabayasi1952, Nelson:1966sp, Yasue:1981wu, Goldstein1987, Baublitz:1988, Wallstrom:1988zf, Budiyono:2017irg, Lindgren:2019tdd},
 for a comprehensive discussion see \cite{Derakhshani:2018ljy}. These approaches are mostly variations of an interpretation called  ``stochastic mechanics'', with diverse justifications for the stochastic behaviour and different perspectives on how to accommodate complex numbers.
 \\
  \vspace{-0cm}
 \\
In this paper, we not only confirm that quantum mechanics \emph{is} a theory of stochastic processes, but we derive it from first principles as \emph{the natural theory of smooth stochastic paths}. These paths are geometrically distinct from the non-differentiable and unphysical paths from Feynman's path integral \cite{Feynman:1948ur}. Instead, they are similar to the paths of de Broglie and Bohm \cite{deBroglie:1928, Bohm:1951xx} in that they are differentiable, but are dissimilar in that they are nondeterministic. Furthermore, their derivation is explicitly local and does not need any ad hoc assumptions, such as the guiding equation and the quantum equilibrium hypothesis. A further difference with Bohmian Mechanics is that the quantum potential is \emph{derived} and is shown to be the kinetic energy of stochastic velocity fluctuations. 
\\ 
 \vspace{-0cm}
\\
We do not assume any prior existence of Lagrangians, actions or equations of motion. Instead, such concepts are derived  from first principles. The defining constraints for a joint transition probability density are solved, implying the natural occurrence of complex numbers, the existence of the density matrix and the Born Rule. Polar decomposition of the transition amplitudes and the density matrix yields general stochastic Hamilton-Jacobi equations. The assumption of smooth paths implies the existence of the Quantum continuity equation \cite{Madelung:1926}, identified as a diffusionless Fokker-Planck equation \cite{Fokker:1914, Planck:1917}.  Smoothness also implies the existence of a stochastic Lagrangian that is similar to its classical counterpart, analogous to what happens in the path integral formalism \cite{Feynman:1948ur}. Restricting to stochastic processes with up to quadratic velocities in the stochastic Lagrangian and neglecting velocity fluctuations yields the classical Hamilton-Jacobi equation \cite{Hamilton:1833, Hamilton:1834, Jacobi:1842} for a minimally coupled \cite{Gell-Mann:1956iqa}  spinless particle. If one does not neglect the stochastic velocity fluctuations, Schr\"{o}dinger's equation \cite{Schrodinger:1926} follows naturally. Surprisingly, we also find direct probabilistic formulas for the action and Lagrangian in terms of advanced/retarded transition probability densities.
\\
 \vspace{-0cm}
\\
But how can a local realist  hidden variable description of Quantum Mechanics be correct? Didn't Bell prove this is impossible \cite{Bell:1964kc}? 
Our derivation reveals that this \emph{is} possible because Bell's formula for local realism is incomplete. Crucial terms that capture stochastic fluctuations of the hidden variables are missing: the state of a system is not a probability density but a generating function of velocity expectation values. Our derivation and interpretation of Quantum Mechanics does therefore not have a measurement problem, but is nondeterministic. While the origins of this nondeterminism is not known, it does suggest the existence of local degrees of freedom external to the particle that make it oscillate stochastically, similar to Einstein and Smoluchowski's analysis of Brownian motion\cite{Einstein:1905:BrownianMotion}.
\pagebreak
\section{General stochastic processes}
We are interested in general stochastic processes for which both  time and position take continuous values on the real line: $\{X_t\}_{t\in \mathbb{R}}$.
Such a system is (over)completely specified by the collection of finite $n$ joint probability densities \cite{vankampen},  
\beq
P(X_{t_n} = x_{n} \text{ and }X_{t_{n-1}} = x_{{n-1}} ... \text{ and }X_{t_0} = x_{0}).
\eeq
Let us study the ``joint transition probability density'' and introduce a simplified notation,
\beq
P(X_{t'} = x' \text{ and } X_t = x) = P(x',t';x, t).
\eeq
%
\subsection{Joint transition probability densities are positive and space-time symmetric}
Probability densities are naturally constrained to be positive,
\beq
P(x',t';x, t) > 0 \qquad   \forall \quad x',t',x,t.
\eeq
Joint transition densities are also  ``space-time symmetric''  by definition because of their logical and relation,
\beq
P(X_{t'} = x' \text{ and } X_t = x) = P(X_t = x \text{ and } X_{t'} = x' ),
\eeq
which implies a second constraint on the simplified form, 
\beq
P(x',t';x, t) = P(x, t;x',t').
\eeq
Positivity and permutation symmetry of space-time pairs are two of Kolmogorov's consistency conditions \cite{vankampen}.
\subsection{Positivity + space-time symmetry $\rightarrow$ generalised Born rule}
Any positive and space-time symmetric function can be written as the symmetric part of a square,
\beq
P(x', t'; x, t) = \tfrac{1}{2}\left(w^2(x',t'; x,t) + w^2(x,t; x',t')\right).
\eeq
In terms of space-time (anti)symmetric parts this gives,
\beq
P(x', t'; x, t) = w_s^2(x', t'; x, t) + w^2_a(x', t'; x, t),
\eeq
which one can view as a generalisation of the Born rule,
\beq
P(x', t'; x, t)= |{\bf w}(x', t'; x,t)|^2.
\eeq
${\bf w}$ is a ``joint transition amplitude'',
\beq
{\bf w}(x', t'; x, t) = w_s(x', t'; x, t) + i w_a(x', t'; x, t),
\eeq
that is ``space-time Hermitian'',
\beq \label{amplitude-spacetime-hermiticity}
{\bf w}^\ddag(x', t'; x, t) \equiv {\bf w}^*(x, t; x', t')  = {\bf w}(x', t'; x, t),
\eeq
%
%
implying that complex conjugation is equal to space-time reversal,
\beq
{\bf w}^*(x', t'; x, t) = {\bf w}(x, t; x', t'). 
\eeq
%
\subsection{Generalised Born rule $\rightarrow$ density matrix}
The generalised Born rule  implies that probabilistic normalisation can be written in quantum mechanical form,
\beq
 \int dx' dx P(x', t'; x,t)  = \int dx'' dx' \delta(x'' - x') \rho(x', t'; x'', t') = \text{tr}(\hat{\rho}) = 1,
\eeq
where the joint transition amplitude is a ``square root'' of the, not necessarily equal time, density matrix,
\beq \label{sqrt}
\rho(x'',t'', x',t') = \int dx \ {\bf w}(x'', t''; x,t){\bf w}^*(x', t'; x,t). 
\eeq
Using space-time Hermiticity of ${\bf w}$, we see that $\rho$ really is a density matrix since it has the defining properties,
\paragraph{(space-time) Hermitian}
%
\beq
(\rho(x'',t'';x',t'))^\ddag = \int dx \ {\bf w}^*(x', t'; x,t) {\bf w}(x'', t''; x,t)= \rho(x'',t''; x',t'),
\eeq
%
\paragraph {trace 1}
%
\beq
\int dx' \rho(x',t', x',t') = \int dx' dx \ |{\bf w}(x', t'; x,t)|^2 = \int dx' P(x',t') = 1,
\eeq
%
\paragraph{positive semi-definite}
%
\beq
\int dx''dx'dx \ \chi(x''){\bf w}(x'', t'; x,t) \chi^*(x'){\bf w}^*(x', t'; x,t)= \int dx \ |\tilde{\chi}(x)|^2 \geq 0.
\eeq
Note that these are not postulated, they simply follow from the existence of a joint transition probability density.

\subsection{Stochastic Hamilton-Jacobi equations}
Since the time associated with the integration variable in \myref{sqrt} does not influence the value of the density matrix, one can conclude that 
the density matrix captures the complete instantaneous state of the stochastic process. If we  do a polar decomposition of the joint transition amplitude, where $r$ is space-time symmetric and $s$ is space-time antisymmetric,
\beq \label{polar:w}
 {\bf w}(x', t;' x, t) = r(x', t'; x, t) e^{i s(x', t'; x, t)},
\eeq
and if we introduce ${\bf w} = {\bf w}( x', t'; x, t)$, it follows that,
\begin{align}
 \tfrac{1}{2i}({\bf w}^*\partial_{x'}  {\bf w} - {\bf w} \partial_{x'}{\bf w}^*) &= P(x', t'; x, t) {\partial}_{x'} s( x', t'; x, t),
 \\
 \tfrac{1}{2i}({\bf w}^*\partial_{\mskip 2mu t'}  {\bf w}\mskip 2mu - {\bf w} \partial_{\mskip 2mu t'}{\bf w}^*) &= P(x', t'; x, t) {\partial}_{t'} s(x', t'; x, t).
\end{align}
Subsequently, there are two sets of wave number and frequency densities, namely the left integrals,
\begin{align} 
\label{stochastic:hamilton:jacobi:x}
&\langle k \rangle_{l} P \equiv \tfrac{1}{2i}\int dx' ( {\bf w}^*\partial_{x'}  {\bf w} - {\bf w} \partial_{x'}{\bf w}^* )  = \int dx' P(x', t'; x, t) {\partial}_{x'} s( x', t'; x, t), 
\\
-&\langle \Omega \rangle_l P \equiv \tfrac{1}{2i}\int dx' ( {\bf w}^*\partial_{t'}  {\bf w} - {\bf w} \partial_{t'}{\bf w}^* )= \int dx' P(x', t'; x, t) {\partial}_{t'} s(x', t'; x, t),
\label{stochastic:hamilton:jacobi:t}
\end{align}
\vspace{-0.4cm}

\noindent and the right integrals that more directly yield $\langle k \rangle = \on{\rightarrow}{\partial}_{\!x} s_\rho $ and $\langle \Omega \rangle = -\on{\rightarrow}{\partial}_{\!t} s_\rho$,
\begin{align} 
\label{stochastic:hamilton:jacobi:x}
&\langle k \rangle P \equiv \tfrac{1}{2i}(\on{\rightarrow}{\partial}_{\!x'}  \rho -  \rho \on{\leftarrow}{\partial}_{\!x'} ) = P \on{\rightarrow}{\partial}_{\!x'} s_\rho = \int dx P(x', t'; x, t) {\partial}_{x'} s( x', t'; x, t), 
\\
-&\langle \Omega \rangle P  \equiv \tfrac{1}{2i}(\on{\rightarrow}{\partial}_{\!t'}  \rho -  \rho \on{\leftarrow}{\partial}_{\!t'} )= P \on{\rightarrow}{\partial}_{\!t'} s_\rho = \int dx P(x', t'; x, t) {\partial}_{t'} s(x', t'; x, t),
\label{stochastic:hamilton:jacobi:t}
\end{align}
where we used an ``element-wise'' polar decomposition of the density matrix into space-time (anti)symmetric parts,
\beq
\rho(x', t'; x, t) = r_\rho(x', t'; x, t)e^{is_\rho(x', t'; x, t)}.
\eeq
The left and right stochastic Hamilton-Jacobi equations have the same ``full'' expectation values,
\begin{align} 
\label{full:stochastic:hamilton:jacobi:x}
&\langle\!\langle k \rangle\!\rangle\, =\, \int \langle k \rangle_{l} P = \,\int \langle k \rangle P \, = \ \int dx' dx P(x', t'; x, t) {\partial}_{x'} s( x', t'; x, t), 
\\
&\langle\!\langle \Omega  \rangle\!\rangle = \int \langle \Omega \rangle_l P = \int\langle \Omega \rangle P = -\!\int\! dx' dx P(x', t'; x, t) {\partial}_{t'} s(x', t'; x, t).
\label{full:stochastic:hamilton:jacobi:t}
\end{align}
Note that we use the following shorthands for ``diagonal operations'',
\begin{align}
\rho = \rho(x, t; x, t) = P(x, t) = P, \qquad 
&\partial_{t}  \rho = \on{\rightarrow}{\partial}_{t}  \rho +  \rho \on{\leftarrow}{\partial}_{t}   = \partial_t P,
\\
\on{\rightarrow}{\partial}_{t}  \rho = \lim_{t'\rightarrow t}\partial_{t'}  \rho(x, t'; x, t), \qquad &\rho \on{\leftarrow}{\partial}_{t} = \lim_{t'\rightarrow t}\partial_{t}  \rho(x, t; x, t'),
\\
\on{\rightarrow}{\partial}_{\!x}  \rho =\! \lim_{x'\!\rightarrow x}\!\partial_{x'}  \rho(x', t; x, t), \qquad &\rho \on{\leftarrow}{\partial}_{\!x} =\! \lim_{x'\rightarrow x}\! \partial_{x}  \rho(x, t; x', t).
\end{align}
%
\section{Smooth stochastic processes}
Let us restrict ourselves to smooth paths, i.e. paths for which all stochastic velocities exist,
\beq \label{stochasticvelocities}
\int d \Delta x (\Delta x)^n P(x + \Delta x, t + \Delta t; x, t)=  (\Delta t)^n \langle v^n \rangle(x, t) P(x, t) + \mathcal{O}((\Delta t)^{n'>n}),
\eeq
which means that the joint transition probability density is strongly peaked in $\Delta x$ for small $\Delta t$.
\subsection{Kramers-Moyal}
The stochastic velocities appear naturally in the small time behaviour of the right hand marginality equation,
\beq \label{marginality}
P(x, t + \Delta t)= \int d \Delta x P(x, t + \Delta t;x - \Delta x, t).
\eeq
Reparametrising $x \rightarrow x + \Delta x - \Delta x$  and expanding in $\Delta x$ gives the standard Kramers-Moyal expansion \cite{Kramers:1940zz, Moyal:1949},
\beq \label{kramersmoyal}
\sum^\infty_{m=0} \tfrac{(\Delta t)^m}{m!} \partial^m_t P(x, t) = \sum^\infty_{m=0} \tfrac{(-1)^m}{m!} \partial^m_{x} \int d\Delta x (\Delta x)^m P(x + \Delta x, t + \Delta t; x, t).
\eeq
Inserting the smooth path condition \myref{stochasticvelocities}, acting with $\partial^n_{\Delta t}$ on both sides and taking the limit $\Delta t \rightarrow 0$ gives
\beq \label{ncontinuity}
\partial^n_{t}  P(x, t)
= 
(-1)^{n}\partial^n_x (\langle v^n \rangle(x,t)P(x, t)). 
\eeq
For $n =1$ we recognise the continuity equation as a diffusionless Fokker-Planck equation where $\langle v\rangle$ is the drift velocity.
\\
\\
We also do a  Kramers-Moyal-like expansion on the right hand Hamilton-Jacobi equations  \myref{stochastic:hamilton:jacobi:x} and \myref{stochastic:hamilton:jacobi:t},
\begin{align} 
P(x,t) \on{\rightarrow}{\partial}_{\!x} s_\rho(x, t; x, t) &= \lim_{\Delta t \rightarrow 0} \sum^\infty_{m=0} \tfrac{(-1)^m}{m!} \partial^m_{x} 
\int d \Delta x (\Delta x)^m P(x + \Delta x, t + \Delta t;x, t)\on{\rightarrow}{\partial}_{\!x} s(x+ \Delta x, t + \Delta t; x, t),
\label{stochhamcobix} 
\\
P(x,t) \on{\rightarrow}{\partial}_{\!t} s_\rho(x, t; x, t)&= \lim_{\Delta t \rightarrow 0} \sum^\infty_{m=0} \tfrac{(-1)^m}{m!} \partial^m_{x} 
\int d \Delta x (\Delta x)^m P(x + \Delta x, t + \Delta t;x, t)\on{\rightarrow}{\partial}_{\!t} s(x+ \Delta x, t + \Delta t; x, t).
\label{stochhamcobit} 
\end{align}
Because the transition probability density is strongly peaked for small $\Delta t$, only small fluctuations contribute,
\beq
s(x + \Delta x, t + \Delta t; x, t) 
=(1 
+ \Delta x \on{\rightarrow}{\partial}_{\!x} 
+ \tfrac{1}{2}(\Delta x)^2 \on{\!\!\rightarrow}{\partial^2}_{\!\!\!\!x} 
+ \tfrac{1}{3!}(\Delta x)^3 \on{\!\!\rightarrow}{\partial^3}_{\!\!\!\!x} 
+... \
)s(x, t + \Delta t; x, t),
\eeq
which leads to well defined smooth stochastic limits if $\on{\!\!\rightarrow}{\partial^n}_{\!\!\!\!x} s(x, t + \Delta t; x, t) = \tfrac{1}{(\Delta t)^{n-1}}l_n(x, t) + \mathcal{O}(\tfrac{1}{\Delta t^{m<n}})$ and therefore,
\beq
s(x + \Delta x, t + \Delta t; x, t) 
= \Delta t\, l_0(x, t) 
+ \Delta x l_1(x, t)
+ \tfrac{1}{2}\tfrac{(\Delta x)^2}{\Delta t} l_2(x,t)
+ \tfrac{1}{3!}\tfrac{(\Delta x)^3}{(\Delta t)^2} l_3(x, t)
+... \ .
\eeq
So when taking the limit $\Delta t \rightarrow 0$, the stochastic Hamilton-Jacobi equations \myref{stochhamcobix} and \myref{stochhamcobit}  reduce to,
\begin{align} 
\on{\rightarrow}{\partial}_x s_\rho & = 
\ \ \, l_1
+\ \,l_2\langle v\rangle
\ \,  + \tfrac{1}{2}l_3\langle v^2\rangle
+ ...\ , \label{wavenumber:velocity}
\\
-\on{\rightarrow}{\partial}_t s_\rho &= 
 - l_0
+ \tfrac{1}{2}l_2\langle v^2\rangle
+  \tfrac{1}{3}l_3\langle v^3\rangle
+ ...\ .\label{freq:velocity}
\end{align}
Because only $m=0$ terms survive the limit, the right and left Hamilton-Jacobi equations become identical,
\begin{align}
\langle k \rangle P = \langle k \rangle_l P=  P\on{\rightarrow}{\partial}_x s_\rho= \lim_{\Delta t \rightarrow 0} \int d \Delta x P(x + \Delta x, t + \Delta t;x, t)\partial_{\Delta x} s(x+ \Delta x, t + \Delta t; x, t),
\\
-\langle \Omega \rangle P = -\langle \Omega \rangle_l P= P\on{\rightarrow}{\partial}_t s_\rho= \lim_{\Delta t \rightarrow 0} \int d \Delta x P(x + \Delta x, t + \Delta t;x, t)\partial_{\Delta t} s(x+ \Delta x, t + \Delta t; x, t).
\end{align}
The above also implies that the stochastic action itself satisfies the following limit,
\beq
\lim_{\Delta t \rightarrow 0}\int d \Delta x P(x + \Delta x, t + \Delta t;x, t) \frac{s(x+ \Delta x, t + \Delta t; x, t)}{\Delta t } \equiv \langle l\rangle(x, t) P(x, t),
\eeq
which defines $\langle l\rangle(x,t)$ as the stochastic Lagrangian,
\beq
\langle l\rangle=  
l_0
+ l_1\langle v\rangle
+ \tfrac{1}{2}l_2\langle v^2\rangle
+  \tfrac{1}{3!}l_3\langle v^3\rangle
+ ... \ .
\eeq
\myref{ncontinuity},\myref{wavenumber:velocity} and \myref{freq:velocity} lead to the following set of diagonal evolution equations for the density matrix,
\begin{align}
\on{\rightarrow}{\partial}_t \rho + \rho \on{\leftarrow}{\partial}_t\ \ &=  - \partial_x (\langle v\rangle\rho), \label{densitymatrix:continuity}
\\
-\tfrac{1}{2}i(\on{\rightarrow}{\partial}_x \rho - \rho \on{\leftarrow}{\partial}_x) &= 
\left( \ \ \, l_1
+\ \,l_2\langle v\rangle
\ \,  + \tfrac{1}{2}l_3\langle v^2\rangle
+ ...\right)\rho, \label{densitymatrix:hamiltonjacobimomentum}
\\
\tfrac{1}{2}i(\on{\rightarrow}{\partial}_t \rho - \rho \on{\leftarrow}{\partial}_t) &= \left( - l_0
+ \tfrac{1}{2}l_2\langle v^2\rangle
+  \tfrac{1}{3}l_3\langle v^3\rangle
+ ... \right)\rho. \label{densitymatrix:hamiltonjacobi}
\end{align}
%
\section{Quadratic stochastic processes}
Processes for which $l_{n>2} = 0$ are particularly interesting as they have an essentially classical momentum equation
\begin{align}
\on{\rightarrow}{\partial}_x s_\rho & = 
l_1
+l_2\langle v\rangle,
\end{align}
derived without using variational principles, equations of motion, Legendre transforms or canonical transformations. Quadratic processes are also interesting because one can solve for the drift velocity,
\beq \label{driftvsol}
\langle v\rangle = \tfrac{1}{l_2}(\on{\rightarrow}{\partial}_x s_\rho - l_1). 
\eeq
%
\subsection{The classical Hamilton-Jacobi equation}
Neglecting the velocity fluctuations $\delta v^2 = \langle v^2\rangle - \langle v \rangle^2$ in \myref{freq:velocity} with $l_{n>2} = 0$ gives a classical Hamilton-Jacobi equation,
\begin{align}
-\on{\rightarrow}{\partial}_t s_\rho&= 
 - l_0
+ \tfrac{1}{2}l_2\langle v\rangle^2, \label{quadraticstochastichamiltonjacobi}
\end{align}
again derived without using familiar classical procedures. Inserting the result for the drift velocity \myref{driftvsol} gives,
\begin{align}
-\on{\rightarrow}{\partial}_t s_\rho&= 
 - l_0
+ \tfrac{1}{2l_2}\big(\on{\rightarrow}{\partial}_x s_\rho -l_1\big)^2, \label{classicalhamiltonjacobi}
\end{align}
which when relabelling to standard  (particle) physics variables $s_\rho(x, t; x, t) = \tfrac{S(x,t; x,t)}{\hbar}$ and
\begin{align}\label{standard:variables}
&-l_0(x,t)= \tfrac{e \varphi(x, t) + V(x, t)}{\hbar} \qquad l_1(x, t) = \tfrac{e A(x, t)}{\hbar} \qquad l_2(x, t) = \tfrac{m(x,t)}{\hbar},
\end{align}
becomes \emph{the} classical Hamilton-Jacobi equation for a spinless particle minimally coupled to an electromagnetic field,
\begin{align}
-\on{\rightarrow}{\partial}_t S&= 
\tfrac{1}{2m}\big(\on{\rightarrow}{\partial}_x S - eA\big)^2
+ e \varphi+ V. \label{classicalhamiltonjacobiequation}
\end{align}
Note that this is actually a slight generalisation of the standard classical Hamilton-Jacobi equation since  $\on{\rightarrow}{\partial}_x S(x, t; x, t)  = \partial_x S(x,t)$
is only true if the density matrix describes a pure state, where $\tfrac{S(x,t)}{\hbar}$ is the phase of the wave function.
\subsection{Madelung's continuity equation}
Inserting the result for the drift velocity \myref{driftvsol} in the continuity equation \myref{densitymatrix:continuity} and using standard variables \myref{standard:variables} gives,
\beq \label{madelungS}
\partial_t P =\partial_x(\langle v\rangle P ) = -\partial_x(\tfrac{1}{m}(\on{\rightarrow}{\partial}_x S - eA) P),
\eeq
 Classically, $P(x,t)$ and $\langle v \rangle(x, t)$ just model the uncertainty stemming from not precisely known initial position and velocity respectively, not randomness of motion. The form of the continuity equation  relevant for deriving quantum mechanics (see \myref{densitymatrix:continuity},\myref{densitymatrix:hamiltonjacobimomentum}) is the following,
\beq \label{madelung:continuity}
(\on{\rightarrow}{\partial}_t \rho + \rho \on{\leftarrow}{\partial}_t) 
= -\partial_x \left(\tfrac{1}{2 l_2}(\on{\rightarrow}{\partial}_{x}  \rho  - \rho  \on{\leftarrow}{\partial}_{x}) - i\tfrac{l_1}{l_2} \rho\right). 
\eeq
%

\section{Repeatable quadratic processes with constant mass}
\subsection{The Schr\"{o}dinger equation}

In the appendix we show that general repeatable stochastic processes evolve according to $i \on{\rightarrow}{\partial}_t \rho = \on{\rightarrow}{\Omega} \rho$.
For quadratric processes with constant $l_2$, $\hat{\Omega}$ is uniquely determined by the continuity equation \myref{madelung:continuity} and the classical Hamilton-Jacobi equation \myref{classicalhamiltonjacobi}. The continuity equation requires the frequency operator to be of the following form,
\beq
\hat{\Omega} = -\tfrac{1}{2 l_2}\partial^2_{x} + i\tfrac{ l_1}{ l_2} \partial_x  + i\tfrac{1}{2 l_2} \partial_x l_1  + \phi.
\eeq
The fact that $i \tfrac{1}{2}(\on{\rightarrow}{\partial}_t \rho - \rho \on{\leftarrow}{\partial}_t ) 
= \tfrac{1}{2}(\on{\rightarrow}{\Omega} \rho +  \rho \on{\leftarrow}{\Omega} ) $  must reduce to \myref{classicalhamiltonjacobi} for $\langle v^2\rangle - \langle v\rangle^2 \rightarrow 0$ implies that $\phi = -l_0 + \tfrac{l^2_1}{2l_2}$ and
\begin{align}
\label{gaugeinvariantfrequencyoperator}
&\hat{\Omega} = -l_0 + \tfrac{1}{2 l_2}(-i\partial_{x} - l_1)^2,
\end{align}
which yields the standard Schr\"{o}dinger equation for minimally coupled spinless particles, 
\beq \label{minimallycoupledschrodinger}
i \hbar \on{\rightarrow}{\partial}_t \rho=  \left(\tfrac{1}{2m}(-i\hbar\on{\rightarrow}{\partial}_{x} - e A)^2 + e\varphi + V\right) \rho.
\eeq
Note that one automatically obtains the Schr\"{o}dinger equation that is covariant under local phase transformations,
\beq
\rho \rightarrow e^{i\chi} \rho, \qquad e\varphi \rightarrow e \varphi - \partial_t \chi, \qquad eA \rightarrow eA - \partial_x \chi.
\eeq
This means that gauge invariance of the Hamiltonian is not a consequence of some secondary principle but is automatically true for quadratic processes with a nonzero linear term in the stochastic Lagrangian.
\\
\\
Note also that we have not used particle physics variables from the start because the derivation is not necessarily about elementary particles. Any repeatable quadratic stochastic process is described by the formalism, including analogue systems for Quantum mechanics where $\hbar$ can be much larger than Planck's constant.  Another reason to do so is that it emphasizes the fact that Planck's constant is just a constant for converting units and scales which implies that it has a similar status as the speed of light $c$.

\subsection{Kinetic energy}
The equality of the stochastic velocity and operator forms of the stochastic Hamilton-Jacobi equation with $\hat{H} = \hbar \hat{\Omega}$
\beq
-P\on{\rightarrow}{\partial}_t S= 
 ( \tfrac{1}{2}m\langle v^2\rangle + e\phi + V)P
=  \tfrac{1}{2}(\on{\rightarrow}{H} \rho +  \rho \on{\leftarrow}{H} ), 
\eeq
gives us the operator form of the kinetic energy,
\beq
P\langle E_{\text{kin}}\rangle  =  \tfrac{1}{2} m \langle v^2 \rangle P = \tfrac{1}{2}( \on{\rightarrow}{H}_{\text{kin}} \rho  +  \rho\on{\!\leftarrow}{H}_{\text{kin}} ),
\eeq
where,
\beq
\hat{H}_{\text{kin}} = \tfrac{1}{2m}(-i\hbar\on{\rightarrow}{\partial}_{x} - e A(x,t))^2. 
\eeq
%
\subsubsection{The velocity operator}
The kinetic energy operator can also be written using velocity operators, 
\begin{align}
&\hat{H}_{\text{kin}} = \tfrac{1}{2} m\hat{v}^2\label{kinop} \quad \text{where}\quad\hat{v} =  \tfrac{1}{m}(-i\hbar\partial_{x} - eA).
\end{align}
Unlike in standard discussions on quantum mechanics, $\hat{v}$ now truly deserves the name ``velocity operator'' as it is directly related to the geometric properties of concrete paths through integrals over the stochastic fluctuations:
\beq \label{squaredvelocity}
\langle\!\langle v^2 \rangle\!\rangle \equiv  \lim_{\Delta t \rightarrow 0}\int dx d\Delta x (\tfrac{\Delta x}{\Delta t})^2 P(x + \Delta x, t + \Delta t; x, t) = \text{tr}( \hat{v}^2 \rho).
\eeq
%
\subsubsection{The Quantum potential}
The stochastic Hamilton-Jacobi equation for finite velocity fluctuations \myref{quadraticstochastichamiltonjacobi} can be written as,
\begin{align}
-\on{\rightarrow}{\partial}_t S&\, = \ \
\tfrac{1}{2} m\langle v\rangle^2 \
+\ \tfrac{1}{2} m (\langle v^2\rangle - \langle v \rangle^2)
+ e \varphi+ V. \label{classicalhamiltonjacobiequation}
\end{align}
But it is also equal to the real part of the Schr\"{o}dinger equation divided by $P(x, t)$,
\begin{align}
-\on{\rightarrow}{\partial}_t S&= 
\tfrac{1}{2m}\big(\on{\rightarrow}{\partial}_x S - eA\big)^2\!\!
- \tfrac{\hbar^2}{2m}\frac{\on{\rightarrow}{\partial^2}_{\!\!\!x} r_\rho}{r_\rho}
+ e \varphi+ V. \label{classicalhamiltonjacobiequation}
\end{align}
This clearly shows that the Quantum potential is the kinetic energy of stochastic velocity fluctuations, 
\beq
V_Q \equiv  -\tfrac{\hbar^2}{2m}\frac{\on{\rightarrow}{\partial^2}_{\!\!\!x} r_\rho}{r_\rho} = \tfrac{1}{2} m (\langle v^2\rangle - \langle v \rangle^2),
\eeq
which resolves the mysteries around its origin and physical interpretation. 
\\ 
\\ 
Since there is no $\hbar$ in the continuity equation $\myref{madelungS}$ we see that the only effect of $\hbar$ on the dynamics of the stochastic process is to set the size of the stochastic velocity fluctuations. So analogue systems with smooth velocity fluctuations that are large correspond to processes with a large effective $\hbar$. It also means that the classical limit, the limit of small stochastic velocity fluctuations and the limit $\hbar \rightarrow 0$ are all the same.
\\
\\
So from our perspective there is no real interpretational gap between quantum mechanics and classical mechanics. For both cases the interpretation of $S$ is the same and the whole physical system is not only described by a  Hamilton-Jacobi equation but also by Madelung's continuity equation. The only difference between classical and quantum mechanics is that because in the classical limit the stochastic velocity fluctuations are neglected, the trajectories become deterministic and there is therefore no dispersion in the Hamilton-Jacobi equation. When keeping the stochastic fluctuations however, there is both uncertainty in the initial conditions as well as ``actual'' uncertainty arising from the nondetermistic motion. The sole term that differentiates quantum from classical mechanics is the kinetic energy of the stochastic velocity fluctuations, which we showed is the quantum potential.

\subsubsection{Negative kinetic energy}
The quantum potential can make the stochastic kinetic energy density negative, but looking at the stochastic integral expression in \myref{squaredvelocity} it would seem that it should be positive as in the classical case. This is just an artefact of the finite $\Delta t$ expression however, the limits for the stochastic velocities should be understood in the following manner,
\beq \label{vn}
 \langle v^n \rangle(x,t) P(x,t) 
 \equiv  \lim_{\Delta t \rightarrow 0}\int d\Delta x (\tfrac{\Delta x}{\Delta t})^n P(x + \Delta x, t + \Delta t; x, t) 
 = \tfrac{1}{n!}\int d\Delta x  (\Delta x)^n\on{\!\!\!\rightarrow}{\partial^n}_{\!\!\!\!t} P(x + \Delta x, t; x, t). 
\eeq
and
\beq \label{vmnotn}
\int d\Delta x  (\Delta x)^{m\neq n}\on{\rightarrow}{\partial^n}_{\!\!\!t} P(x + \Delta x, t; x, t) = 0. 
\eeq
From this it is clear that even stochastic velocities $\langle v^{2m} \rangle$ do not have to be positive for $m>0$ since
\beq \label{vn:dtnp} 
\on{\!\!\!\!\!\!\rightarrow}{\partial^{2m}}_{\!\!\!\!\!\!\!\!t}\ \ P(x + \Delta x, t; x, t) \ngeqslant 0 \quad \forall\quad x, \Delta x.
\eeq
So, crucially, stochastic velocities are not standard expectation values in the following sense,
\beq \label{vn:neqp}
 \langle v^n \rangle (x, t)P(x, t) \neq \int d v v^n P_v(x, v, t) \qquad P_v(x, v, t) > 0.
\eeq
Therefore the kinetic energy density and its expectation value are generally not positive everywhere.
This stochastic geometric  insight into negative kinetic energy could be important for a deeper understanding of Quantum tunneling phenomena.
\subsection{Instantaneous state}
\myref{vn} and \myref{vmnotn} imply that the joint transition density is a strongly peaked distribution in the following sense,
\beq
\on{\!\!\!\rightarrow}{\partial^n}_{\!\!\!\!t} P(x + \Delta x, t; x, t) = (-1)^n \delta^{(n)}(\Delta x)\langle v^n \rangle(x, t) P(x, t) . 
\eeq
Heuristically this means that the transition probability density has the following expansion for small $\Delta t$,
\beq
P(x + \Delta x, t + \Delta t; x, t)
 = \left(\delta(\Delta x) - \Delta t\delta'(\Delta x)\langle v \rangle(x, t)  + \tfrac{1}{2}(\Delta t)^2 \delta''(\Delta x)\langle v^2 \rangle(x, t)+ ...\right)P(x, t) . 
\eeq
Concretely, it means that the instantaneous state for stochastic transitions is captured by a generating function, 
\beq \label{zeta}
\zeta_v(x, \alpha, t) 
= \int d\Delta x \sum_{n=0}^{\infty}\tfrac{1}{n!}(i\alpha \Delta x)^n (-1)^n\delta^{(n)}(\Delta x) \langle v^n \rangle(x, t) P(x, t)
= \sum_{n=0}^{\infty}\tfrac{1}{n!}(i\alpha)^n \langle v^n \rangle(x, t) P(x, t) .
\eeq
Note that while classically  only $\langle v^0\rangle$ and $\langle v^1\rangle$ are relevant, the stochastic state needs all $\langle v^n \rangle$ because $\langle v^n \rangle \neq \langle v \rangle^n$.
\pagebreak
\section{Stochastic process of a free particle}
\subsection{All order velocity operators and probabilistic interpretation of the density matrix} 
To drive home the point that Quantum Mechanics really is a theory about smooth stochastic paths, it is instructive to check that the velocity expectation values are indeed related to arbitrary powers of the momentum operator. 
Let us therefore compute the stochastic fluctuations of a free particle via a generating function method,
\beq
\langle (\Delta x)^n \rangle(x, t, \Delta t) = \lim_{q \rightarrow 0 }(-i\hbar{\partial_q})^n 
\langle e^{\frac{i}{\hbar} q\Delta x} \rangle(x, t, \Delta t)
\eeq
where the generating function is the following stochastic integral,
\beq
 \langle e^{\frac{i}{\hbar} q\Delta x}\rangle(x, t, \Delta t)  =\int d\Delta x
 e^{\frac{i}{\hbar}q \Delta x}  | {\bf w}(x+\Delta x, t+\Delta t;x , t) |^2.
\eeq
The free particle time evolution operator in position space (the propagator) is given by
\beq
U(x + \delta x, t + \Delta t; x, t) 
= \sqrt{\frac{m}{2\pi i \hbar \Delta t}} e^{\frac{i}{\hbar} \Delta t \frac{1}{2} m \left(\frac{\delta x}{\Delta t}\right)^2},
\eeq
which is the solution to the free particle Schr\"{o}dinger equation with initial condition,
\beq
\lim_{\Delta t \rightarrow 0}U(x + \delta x, t + \Delta t; x, t)  = \delta(\delta x).
\eeq
Inserting this into to the generating function gives,
\beq
\langle e^{\frac{i}{\hbar} q\Delta x} \rangle =
\int d y
 e^{\frac{i}{\hbar}q (x - y)} 
{\bf w}(y + \tfrac{1}{2}\tfrac{q}{m}\Delta t, t; x, t)
{\bf w}^*(y - \tfrac{1}{2} \tfrac{q}{m}\Delta t, t; x, t).
\eeq
Expanding in the generating variable $q$ gives,
\beq
\langle e^{\frac{i}{\hbar} q\Delta x} \rangle =
\sum_{k=0}^{\infty}\tfrac{1}{k!}(\tfrac{q}{m}\Delta t)^k\partial^k_{\delta x}\int d y
{\bf w}(y + \tfrac{1}{2}\delta x, t; x, t)
{\bf w}^*(y - \tfrac{1}{2} \delta x, t; x, t)\Big|_{\delta x = 0} 
+ O(q^{l>k} (\Delta t)^k) ).
\eeq
This implies the following expression for the stochastic velocities,
\beq \label{wignermoyalvelocityoperators}
\langle v^n\rangle(x, t) P(x, t)
= \lim_{\Delta t \rightarrow 0}\tfrac{1}{n!} \partial^n_{\Delta t}\langle (\Delta x)^n \rangle 
= \left(\frac{- i\hbar \partial_{\delta x}}{m}\right)^n \int d y
{\bf w}(y + \tfrac{1}{2}\delta x, t; x, t)
{\bf w}^*(y - \tfrac{1}{2} \delta x, t; x, t)\Big|_{\delta x = 0} 
\eeq
where we recognise the bidirectional momentum operators, familiar from the Wigner function formalism. 
This also means that the left integral of the transition amplitudes is a generating function for the stochastic velocities,
\beq
\int d y
{\bf w}(y + \tfrac{1}{2}\delta x, t; x, t)
{\bf w}^*(y - \tfrac{1}{2} \delta x, t; x, t) = \sum_{n=0}^\infty \tfrac{1}{n!} (\tfrac{i m}{\hbar}\delta x)^n \langle v^n\rangle(x, t) P(x, t)
\eeq
where on the right hand side we recognise \myref{zeta} in the form $\zeta(x, \alpha =\tfrac{m}{\hbar} \delta x)$ yielding,
\beq
\sum_{n=0}^{\infty}\tfrac{1}{n!} (i\alpha)^n\int d\Delta x (\Delta x)^n \on{\!\!\!\rightarrow}{\partial^n}_{\!\!\!\!t} P(x + \Delta x, t; x, t) =
\int d y
{\bf w}(y + \tfrac{1}{2}\tfrac{\hbar}{m} \alpha, t; x, t)
{\bf w}^*(y - \tfrac{1}{2} \tfrac{\hbar}{m} \alpha, t; x, t) . 
\eeq
This means that for any $t$ the equal-time transition probability can be reconstructed as follows,
\beq
\on{\!\!\!\rightarrow}{\partial^n}_{\!\!\!\!t} P(x + \Delta x, t; x, t) = (-1)^n\delta^{(n)}(\Delta x) \int d y (-i \partial_\alpha)^n 
{\bf w}(y + \tfrac{1}{2}\tfrac{\hbar}{m} \alpha, t; x, t)
{\bf w}^*(y - \tfrac{1}{2} \tfrac{\hbar}{m} \alpha, t; x, t). 
\eeq
If we also integrate over $x$, we see that the density matrix is the generating function for velocity expectation values,
\beq \label{densitymatrixisgeneratingfunction}
\int d y
\rho(y + \tfrac{1}{2}\tfrac{\hbar}{m} \alpha, t; y - \tfrac{1}{2}\tfrac{\hbar}{m} \alpha, t) 
= \sum_{n=0}^{\infty}\tfrac{1}{n!} (i\alpha)^n\int dx d\Delta x   (\Delta x)^n \on{\!\!\!\rightarrow}{\partial^n}_{\!\!\!\!t} P(x + \Delta x, t; x, t).
\eeq
This means that the density matrix can be completely understood in terms of the geometry of the stochastic paths.
The reverse is also true: the generating function of velocities describes the complete instantaneous state of the process.
\\
\\
Partial integration in \myref{wignermoyalvelocityoperators} yields the left hand momentum density in terms of standard momentum operators,
\beq
m^n \langle v^n\rangle(x, t) P(x, t)
= \Re \int d y
{\bf w}^*(y, t; x, t) (- i\hbar \partial_{y})^n {\bf w}(y, t; x, t),
\eeq
implying $\langle v^n\rangle(x, t) = \left(\tfrac{\hbar}{m}\right)^n\langle k^n \rangle_l(x, t)$ and a natural expression for the full expectation values,
\beq
m^n\langle\!\langle v^n \rangle\!\rangle 
= \tfrac{1}{n!} \int dx d\Delta x (m\Delta x)^n \on{\!\!\!\rightarrow}{\partial^n}_{\!\!\!\!t} P(x + \Delta x, t; x, t)  
= \text{tr} ((- i\hbar \on{\rightarrow}{\partial}_{y})^n \rho).
\eeq
%

\vspace{-0cm}
\section{Probabilistic meaning of the action and Lagrangian} \label{probmeaning}
\vspace{-.1cm}
So far, the stochastic action and Lagrangian are phase variables and therefore their \emph{direct} probabilistic interpretation is unclear.
In this section we show that they are in fact properties of the retarded transition probability density.
\vspace{-.2cm}
\subsection{Definitions}
\vspace{-.1cm}

We start by decomposing the joint probability density and wave amplitude into retarded and advanced parts,
\begin{align}
P(x', t'; x, t) = P(x', t' > x, t) + P(x', t' < x, t),
\\
w(x', t'; x, t) = w(x', t' > x, t) + w(x', t' < x, t),
\end{align}
where $P(x', t' > x, t)$ and $P(x', t' < x, t)$ are the retarded and advanced transition probabilty densities respectively,
\begin{align}
P(x', t' > x, t) = w^2(x',t' > x,t) \quad\text{where}  \quad P(x', t' > x, t) = 0\quad \text{and} \quad w(x', t' > x, t) = 0\quad \forall \quad t' < t,
\\
P(x', t' < x, t) = w^2(x',t' < x,t)\quad \text{where} \quad P(x', t' < x, t) = 0\quad \text{and} \quad w(x', t' < x, t) = 0\quad \forall \quad t' > t.
\end{align}
Because the joint transition probability density is space-time symmetric, the advanced transition probability density is the space-time reverse of its retarded counterpart, analogous to advanced and retarded propagators,
\beq
P(x', t' < x, t) = P(x, t > x', t').
\eeq
We decompose the retarded density into space-time (anti)symmetric parts as follows,
\beq
 P(x', t' > x, t) = P_s(x', t'; x, t) + f_a(x', t'; x, t), 
\eeq
where it should be noted that $P(x', t' > x, t)$ and $P_s(x', t'; x, t)$ are probability densities while $f_a(x', t'; x, t)$ is not,
\begin{align}
 P_s(x, t ; x', t' )&= \tfrac{1}{2}(P(x', t' > x, t) + P(x,' t' < x, t)) = \tfrac{1}{2} P(x', t'; x, t) ,
 \\
 f_a(x, t; x', t') &=\tfrac{1}{2}(P(x', t' > x, t) - P(x', t' < x, t)).
\end{align}
If we also introduce the space-time (anti)symmetric parts of the retarded wave amplitude, 
\begin{align}
w_s(x', t' > x, t) = \tfrac{1}{2}(w(x', t' > x, t) + w(x, t > x', t')),
\\
w_a(x', t' > x, t) =\tfrac{1}{2}(w(x', t' > x, t) - w(x, t > x', t')),
\end{align}
we conclude that,
\begin{align}
P(x', t' > x, t) &= w_s^2(x',t' > x,t) + w_a^2(x',t' > x,t) + 2w_s(x',t' > x,t)w_a(x',t'> x,t).
\\
\tfrac{1}{2}P(x', t' ; x, t) &= w_s^2(x',t'> x,t) + w_a^2(x',t'> x,t),
\\
f_a(x', t' ; x, t)&= 2w_s(x',t' > x,t)w_a(x',t' > x,t),
\end{align}
and that the transition amplitudes from the previous sections can be identified with the retarded wave amplitudes,
\begin{align}
w_s(x',t' ; x,t) =  \sqrt{2}w_s(x',t' > x,t),
\\
w_a(x',t' ; x,t) =  \sqrt{2}w_a(x',t' > x,t).
\end{align}
Therefore, the antisymmetric retarded transition density can be expressed in terms of the original amplitudes,
\beq \label{antisymmetricP:oldw}
f_a(x', t' ; x, t) 
= w_s(x',t'; x,t)w_a(x',t'; x,t).
\eeq
\vspace{-.7cm}
\subsection{Results}
\vspace{-0cm}
Using the polar decomposition \myref{polar:w} in \myref{antisymmetricP:oldw} yields the direct probabilistic formula for the stochastic action,
\beq \label{probabilistic:action}
P(x',t'; x,t) \sin (2s(x',t'; x,t))= P(x', t' > x, t) - P(x', t' < x, t).
\eeq
 Dividing by $t'- t$ and taking the equal time limit gives the direct probabilistic formula for the stochastic Lagrangian,
\beq
\langle l \rangle(x, t)P(x, t)= \tfrac{1}{2}\lim_{t' \rightarrow t}\int dx' \frac{(P(x', t > x, t) - P(x', t < x, t))}{t'-t}.
\vspace{-.3cm}
\eeq
So the symmetric part of the retarded joint probability density is related to the standard transition probability as used in our derivation of the Schr\"{o}dinger equation and the antisymmetric part is related the action and Lagrangian. 
\section{Non-Bell local realism} 
In the following we write $\lambda$ instead of $x$ to adhere to conventions used in Bell inequality discussions \cite{Bell:1964kc} as it implies more generality.
To recap, expectation values for stochastic observables cannot generally be written as follows,
\beq \label{hiddenexpectationvalue}
\langle\!\langle A \rangle\!\rangle \neq \int d\lambda A(\lambda) P(\lambda)
\eeq
because it neglects dynamical information of the hidden variables, instead the stochastic nature requires,
\beq
\langle\!\langle A \rangle\!\rangle \nonumber
= \lim_{\Delta t \rightarrow 0}\int d\lambda  d\Delta \lambda A(\lambda, \tfrac{\Delta \lambda}{\Delta t}) P(\lambda, \Delta \lambda, \Delta t)
= \lim_{\Delta t \rightarrow 0}\int d\lambda  d\Delta \lambda 
\sum^\infty_{n=0} \tfrac{1}{n!} A_n (\lambda) \left(\tfrac{\Delta \lambda}{\Delta t}\right)^n
\sum^\infty_{m=0} \tfrac{1}{m!}(\Delta t)^m\partial^m_t P(\lambda, \Delta \lambda, \Delta t).
\eeq
Smoothness requires that only the $m =n$ terms survive and since these were shown to be proportional to quantum momentum operators 
one is led to the Quantum expression for expectation values, 
\beq
\langle\!\langle A \rangle\!\rangle = \int d\lambda \hat{A}(\lambda, \on{\rightarrow}{\partial}_{\lambda}) \rho(\lambda;  \lambda),
\eeq
Crucial here are \myref{vn}, \myref{vn:dtnp} and \myref{vn:neqp}. Implicit in Bell's hidden variable formula for the expectation value \myref{hiddenexpectationvalue} is that the instantaneous state is a probability distribution but from \myref{vn},\myref{vn:dtnp} and \myref{vn:neqp} we see that that is not the case for smooth stochastic paths. Instead, as illustrated by our free particle calculations, the instantaneous state is given by a generating function of velocity expectation values, i.e. the density matrix \myref{densitymatrixisgeneratingfunction}. 
So while at finite $\Delta t$ the system is naturally described probabilistically by the transition probability, in the instantaneous limit   $\Delta t \rightarrow 0$ the natural state description is the velocity generating function.
The diagonal part of this state function still describes a probability distribution, but the off-diagonal terms are non-positive distributions that capture the velocity expectation values.
Consequently, also for a two particle system the instantaneous state is not a probability distribution, contrary to what is assumed in Bell inequality derivations, and we therefore do not have standard probabilistic expectation values,
\beq
\langle\!\langle A B \rangle\!\rangle \neq \int d\lambda_a d\lambda_b A(\lambda_a) B(\lambda_b) P(\lambda_a, \lambda_b)
\eeq
which is the two sided version of Bell's local realism formula \cite{Bell:1964kc}. 
Again, the expectation values not only depend on the distribution of the values of the stochastic hidden variables but also on their velocity expectation values, 
\beq
\langle\!\langle A B\rangle\!\rangle = \lim_{\Delta t \rightarrow 0}
\int d\lambda_a  d\Delta \lambda_a d\lambda_b  d\Delta \lambda_b
A(\lambda_a, \tfrac{\Delta \lambda_a}{\Delta t})B(\lambda_b, \tfrac{\Delta \lambda_b}{\Delta t}) 
P(\lambda_a, \Delta \lambda_a; \lambda_b, \Delta \lambda_b; \Delta t),
\eeq
which exactly as in the single particle case leads to the proper quantum mechanical expression,
\beq
\langle\!\langle A B\rangle\!\rangle =
\int d\lambda_a  d\lambda_b
\hat{A}(\lambda_a, \on{\rightarrow}{\partial}_{\lambda_a})
\hat{B}(\lambda_b, \on{\rightarrow}{\partial}_{\lambda_b}) 
\rho(\lambda_a, \lambda_b; \lambda_a, \lambda_b),
\eeq
which is well known to lead to violations of Bell's inequalities.\\ 
\section{Conclusion} 
We have derived Quantum Mechanics  as the natural theory for describing stochastic processes with smooth paths.
It is a ``nondeterministic locally realistic hidden variable theory'' that violates Bell's inequalities because Bell's local realism formula misses essential terms coming from the stochastic fluctuations of the local hidden variables.
\\
\\
This resolves many philosophical problems of Quantum theory. In our stochastic formulation of Quantum \emph{Mechanics}, a quantum is an ``element of reality'' (i.e. it is ontic) because it is a particle that is always at a specific location. 
The density matrix is not an ``element of reality'' as it encodes the information of the joint transition probability. The density matrix is therefore epistemic or doxastic, depending on ones philosophical view of probabilities.
This means that there is no collapse of the wavefunction and no measurement problem, a measurement is just an update of knowledge/belief. From an interpretational perspective the ``spreading of the wave function'' is exactly analogous to taking a position on the stock market. If you take a position at some time $t_0$ and don't check in on the value of your stock for the next few days, you know it has a real value, but your estimation for that value has a certain probability distribution you assign to it based on past experience. Once you check in on the actual value, you update your knowledge by checking its real value. One could say the probability distribution has collapsed but since your probability estimate for the value was not an element of reality to begin with such a statement seems overly baroque. Of course a practical distinction between the stock market and quantum mechanics, irrelevant for the above argument, is that the stock market is diffusive while quantum mechanics is not.
While such a psi-epistemic/doxastic but nevertheless realist interpretation has been appealing to many, without a good explanation of how super-Bell correlations can arise, such a position is difficult to argue convincingly. From our stochastic derivation it is clear however, that the infinite differentiability of the paths has the capacity to store more local statistical information than one would naively expect, which leads to super-Bell correlations and might be seen as a foundational feature that enables Quantum computing.
\\
\\
Measurement does generally induce ``finite and uncontrollable interaction''  \cite{Bohr:1935af} between the measurement apparatus and the particle in the sense that ``coherence is lost'' because a measurement apparatus will disturb the extremely tiny and fragile stochastic oscillations of microscopic particles. Physically this is mostly important when describing consecutive measurements on the same physical system such as when one tries to measure a particle going through one of the slits in the double slit experiment.  Measurements are active interventions that change the stochastic oscillations of the particle that destroy the interference pattern. Conceptually however, this point does not play a crucial role in our formalism when discussing ensemble results of ``non-consecutive'' measurements of identically prepared systems. There are no ``two sets of rules'' where one is the quantum dynamics as described by the Schr\"{o}dinger equation and the other the theory of measurement.  There is no need for special roles for observers and measurements at all. Quantum mechanical expectation values simply turn out be equivalent to the natural stochastic expectation values, fully inline with and even used to derive the Schr\"{o}dinger equation.  To properly account for consecutive measurements, one ``simply'' needs to also model the measurement apparatus as a stochastic system and its interaction with the stochastic particle. Such modelling is not necessary however to derive the rule for taking standard Quantum expectation values of non consecutive measurements.
\\
\\
``Wave particle duality'' means that a quantum has both particle and wave-like characteristics. In our stochastic description of quantum mechanics the quantum \emph{is} a particle that \emph{behaves} like a wave because it does not move along the ``straight'' paths of minimal action but smoothly and randomly \emph{oscillates} around such classical paths.  These oscillations around the classical paths are the wave-like behaviour. In particular \myref{wavenumber:velocity} and \myref{freq:velocity} are the stochastic incarnations of the de Broglie relations \cite{deBroglie:1925cca}. For a free particle these reduce quite beautifully to
\begin{align}
\hbar\langle k \rangle(x, t)& = 
m\langle v\rangle(x, t),
\\
\hbar\langle \Omega \rangle(x, t)&= 
\tfrac{1}{2}m\langle v^2\rangle(x,t).
\end{align}
The theory is complete in a statistical sense since the density matrix at one time is enough to predict what the probabilistic state will be at a later time, and the density matrix is all one needs to compute all statistical observables of a particle.
The theory is incomplete \cite{Einstein:1935rr} in the sense that the density matrix does not describe the ontological state of reality. Quite oppositely, it describes our incomplete knowledge of other underlying ontological degrees of freedom. It suggests that the position of the particle is ontological and that there are more microscopic degrees of freedom external to the particle, which cause it to move along stochastic trajectories.  Our theory does not explicitly describe the properties of such hypothetical external ontological degrees of freedom but one could argue that they are implicitly modelled in a statistical sense by the values of the $l_n$ variables. A different potential for instance implies a different local behaviour of those ``sub-quantum'' degrees of freedom which means that the particles will be ``guided'' along different trajectories. In case of the double slit experiment for instance, this means that the environment in which the particle moves is genuinely different between one or two open slits. So the stochastic and Bohmian pictures are similar in the sense that external degrees of freedom seem to guide the particle how to move, but they are very different in the sense that in the stochastic picture there is no extra guiding equation, no need for a quantum equilibrium hypothesis and the quantum potential is not a mysteriously given quantity but is derived as the kinetic energy of the stochastic velocity fluctuations. Most importantly however, the stochastic picture not only offers an interpretation but also a derivation of Quantum Mechanics as we have shown above.
\\
\\
Note that our identification of Quantum mechanics as a nondeterministic theory does not necessarily mean that nature is fundamentally nondeterministic. Quantum mechanics could be an effective statistical description of fundamentally deterministic degrees of  freedom or there could be fundamental nondeterminism, our description is agnostic about this.
\\
\\
A deep and surprising insight that follows from our work in section \myref{probmeaning} is that the action and Lagrangian have a direct probabilistic origin as they are properties of the retarded transition probability. Concretely, specifying a physical system by its Lagrangian is in fact the same as specifying moments of the space-time antisymmetric part of the retarded transition probability density.
\\
\\
While our discussion was done for non-relativistic spinless particles, we are confident that this method will generalise well to relativistic systems with sub-structure such as particles with spin, fields, strings etc.
An important reason for this is that from our point of view, the process of quantization is nothing but finding a consistent stochastic theory with smooth fluctuations that reduces to the desired classical theory at scales larger than the oscillation scale. 
Another reason is that the Schr\"{o}dinger equation plays a much more central role in relativistic quantum field theory than typical treatments make it seem.  The Fock space of relativistic quantum field theory is most often built from positive/negative frequency parts of the relativistic fields, a fact that is often hidden by mainly discussing the momentum space representation, but these positive frequency fields satisfy the square root form of the relativistic Schr\"{o}dinger equation. If one takes a string theory-like first quantisation approach using worldines, a covariant form of the Schr\"{o}dinger equation appears in the form of the St\"{u}ckelberg equation \cite{Stueckelberg:1941rg} where the evolution parameter is the proper time of the particle. This equation plays an important role in the Fock-Schwinger proper time method \cite{Itzykson:1980rh}.  The most important appearance of the Schr\"{o}dinger equation in relativistic quantum field theory however, is the so-called functional Schr\"{o}dinger equation which deals with fields as the fluctuating variables instead of particle positions \cite{hatfield:1992}. This equation does not just play a central role in the functional Schr\"{o}dinger formulation of quantum field theory but since the path integral is the time evolution operator corresponding to the functional Schr\"{o}dinger Hamiltonian in field space, the famous path integral itself also satisfies the functional Schr\"{o}dinger equation.  From our point of view, the fundamental object for relativistic quantum field theory is in that case the transition probability from field configuration to field configuration and the functional Schr\"{o}dinger equation follows from the same recipe as we used here to derive the Schr\"{o}dinger equation for a non-relativistic particle. Our formalism does suggest  however an even more natural (over)complete covariant specification of a relativistic quantum field theory in terms of the joint stochastic probability densities for local field values, similar to the treatment of stochastic processes,
\beq \label{fieldprobs}
P(\Phi_{x^\mu_n} = \phi_{n} \text{ and }\Phi_{x^\mu_{n-1}} = \phi_{{n-1}} ... \text{ and }\Phi_{x^\mu_0} = \phi_{0})
=
P(\phi_n, x^\mu_n\,;\, \phi_{n-1}, x^\mu_{n-1}\,; ...\, ;\,\phi_0, x^\mu_0)
.
\eeq
The quantum nature then follows from solving the positivity and permutation-of-pairs symmetry, exactly as we did above for a stochastic process. Also the smoothness condition used above for stochastic paths carries over naturally to the smoothness of stochastic fields. Note that the field expectation values corresponding to the joint probability densities are analogous to Schwinger-Keldysh/in-in expectation values, they do not in general relate to in-out expectation values that are most often treated in textbooks. This is because most textbooks on relativistic quantum field theory focus on computing scattering amplitudes instead of evolution from initial data. Another way to see this, is that expectation values for real fields using \myref{fieldprobs} are manifestly real, as are Schwinger-Keldysh expectation values, whereas in-out expectation values are generally complex valued, even for real fields.

\section*{Acknowledgements}
Special thanks to Floris Hermsen for checking the calculations of this manuscript.
I would also like to thank my colleagues at Owlin for their understanding and for their moral and practical support during the finalisation of this paper and my parents for standing by me all these years and   for playing a large role in enabling me to work on the deeper questions in physics. 

\appendix
\section{Repeatable stochastic processes}
To support the derivation of Quantum Mechanics as the the natural theory of smooth stochastic paths we need to discuss general ``repeatable'', not necessarily smooth, stochastic processes. This section is essentially a recap of generally known Quantum formalism from our stochastic point of view, for systems that are described by the time evolution of a density matrix.

\subsection{Time evolution of general stochastic processes}
Since the time evolution for general stochastic processes has been identified with the time evolution of a general density matrix, one can also describe the time evolution by a nonlinear map that takes an initial density matrix to a density matrix at some later instant \cite{Milz:2018, Preskill:2018},
\beq \label{evolution:map}
\rho_{t'} = \mathcal{E}(\rho_t, t', t), 
\eeq
where the map is required to be continuous and differentiable in time on general physical grounds,
\begin{align}
\lim_{t' \rightarrow t} \mathcal{E}(\rho_t, t', t) &= \rho_t,
\\
\lim_{t' \rightarrow t} \partial_{t'}\mathcal{E}(\rho_t, t', t) &= \mathcal{L}(\rho_t, t)
\end{align}
and should preserve the defining properties of the density matrix as derived above,
\begin{align}
\mathcal{E}(\rho)^\ddag &= \mathcal{E}(\rho),\\ 
\text{tr}\mathcal{E}(\rho) &= 1,\\
\mathcal{E}(\rho) &> 0.
\end{align}
%
\subsection{Time evolution of repeatable stochastic processes}
We define repeatability as the physical requirement that identically prepared experimental settings should lead to statistically identical outcomes. This implies that  statistical expectation values can be understood as averages of identically prepared systems. 
Repeatability encapsulates time-translation invariance in the sense that if one repeats an experiment on the same location at a different time the stochastic particle should leave no memory of the previous experiment.
It encapsulates spatial translational invariance in the sense that if one does multiple experiments in parallel at different locations but similar times, the different experiments cannot non-locally influence each other. This physical requirement that a stochastic process should be repeatable can be translated into the mathematical constraint that the time evolution map of the density matrix \myref{evolution:map} should be linear,
\begin{align} 
\mathcal{E}( p_1 \rho_1 + p_2 \rho_2) = p_1\mathcal{E}(\rho_1) + p_2 \mathcal{E}(\rho_2).
\end{align}
From our stochastic point of view and the insight that repeatability is the physical requirement behind linearity, it is clear that linearity does \emph{not} hold for generic processes. 
An example of a stochastic process that does not satisfy the repeatability requirement is a rider on a dirt bike going around a muddy track.  The trajectory of the dirt bike riding laps around the track is described by a stochastic process. The different laps that the biker takes cannot be seen as independent experiments however since each lap the bike changes the conditions for the next lap because of the tracks it leaves behind. This counterexample shows that linearity is not a feature that must be true by itself but is a genuine physical statement about the type of physical process one is describing. One can of course make the stochastic system linear again by randomly preparing, in a statistically independent manner, many tracks and study the ensemble of those tracks. This is not what one means by identically prepared systems in quantum mechanics however. Typically one envisages situations such as the double slit experiment where particles with the same statistical initial state are sent through the same experimental setup which is analogous to a dirt bike going around the same lap multiple times. In this context, the repeatability constraint is essentially time-translation invariance and physically implies the constraint that each particle going through the slits experiences no statistically discernible effect from the particles than went through the apparatus before it.

It is well known in the literature that linearity of time evolution for the density matrix of a closed system implies unitary evolution but to make our discussion self contained we sketch the derivation below. 
\\
\\
The density matrix is a positive definite Hermitian linear operator and can therefore be diagonalised, 
\beq
\rho= \sum_{n} \langle x' | \tilde{\psi}_n \rangle \langle \tilde{\psi}_n | x \rangle = \sum_{n} c_n \langle x' | \psi_n \rangle \langle \psi_n | x \rangle, 
\eeq
where the $c_n = \langle \tilde{\psi} | \tilde{\psi} \rangle = |\tilde{\psi}_n|^2$ are the positive real eigenvalues that sum to one: $\sum_{n=0}^\infty c_n = 1$. 
They represent the squared norms  of unnormalised basis functions $\tilde{\psi}_n(x,t)$, which for a stochastic particle on the real line can be chosen to be Hermite functions. 
General time evolution can thus be described by nonlinear maps for each basis vector,  
\beq
\psi_{n} = \mathcal{U}_n(\psi_{n}).
\eeq
where similar to $\mathcal{E} $, the map is required to be continuous and differentiable in time on general physical grounds
\begin{align}
\lim_{t' \rightarrow t} \mathcal{U}_n(\psi_{nt}, t', t) &= \psi_{nt},
\\
\lim_{t' \rightarrow t} \partial_{t'}\mathcal{U}_n(\psi_{nt}, t', t) &= \Omega(\rho_t, t),
\end{align}
which implies the following form for the evolution map of the density matrix,
\beq 
\mathcal{E}(\sum_{n} c_n \hat{\rho}_n) = \sum_{n} |\mathcal{U}_n(\tilde{\psi}_n)|^2 \mathcal{U}_n(\psi_n)\mathcal{U}_n(\psi_n)^*.
\eeq
$\mathcal{E}$ becomes a linear operator if all $\mathcal{U}_n$'s are the same unitary linear operator $\hat{U}$,
\beq 
\hat{\mathcal{E}}\sum_{n} c_n \hat{\rho}_n = \sum_{n}c_n\hat{U}\psi_n \psi^*_n\hat{U}^\dag.
\eeq
Therefore, repeatable stochastic processes on the real line have a density matrix that evolves unitarily,
\beq
\rho_{t'} = \rho_{t'\!,t'} = \hat{U}\rho_{t}\hat{U}^\dag,
\eeq
and therefore,
\beq
i\on{\rightarrow}{\partial}_{t}\rho =  \hat{\Omega}_{t} \rho.
\eeq

\end{document}